\documentclass[conference]{IEEEtran}
\IEEEoverridecommandlockouts
\usepackage{cite}
\usepackage{amsmath,amssymb,amsfonts}
\usepackage{algorithmic}
\usepackage{graphicx}
\usepackage{textcomp}
\usepackage{xcolor}

\usepackage{booktabs}
\usepackage{tabulary}
\usepackage[bookmarks=false]{hyperref}  
\usepackage{amsthm}
\usepackage{stfloats}
\usepackage{subfig}  
\usepackage[super]{nth}
\theoremstyle{plain}
\newtheorem{thm}{Theorem}[]
\theoremstyle{definition}
\newtheorem{defn}[thm]{Definition}

\def\BibTeX{{\rm B\kern-.05em{\sc i\kern-.025em b}\kern-.08em
    T\kern-.1667em\lower.7ex\hbox{E}\kern-.125emX}}
\begin{document}

\title{An Evaluation of Bitcoin Address Classification based on Transaction History Summarization\\
}

\author{\IEEEauthorblockN{Yu-Jing Lin\textsuperscript{*}, Po-Wei Wu\textsuperscript{*}, Cheng-Han Hsu\textsuperscript{*}, I-Ping Tu\textsuperscript{$\dagger$}, and Shih-wei Liao\textsuperscript{*}}
\IEEEauthorblockA{\textsuperscript{*} \textit{Department of Computer Science, National Taiwan University, Taiwan} \\
\textsuperscript{$\dagger$} \textit{Institute of Statistical Science, Academia Sinica, Taiwan} \\
r06922068@ntu.edu.tw}
}

\IEEEoverridecommandlockouts
\IEEEpubid{\makebox[\columnwidth]{978-1-7281-1328-9/19/\$31.00~\copyright2019 IEEE \hfill} \hspace{\columnsep}\makebox[\columnwidth]{ }}

\maketitle

\IEEEpubidadjcol

\begin{abstract}
Bitcoin is a cryptocurrency that features a distributed, decentralized and trustworthy mechanism, which has made Bitcoin a popular global transaction platform. The transaction efficiency among nations and the privacy benefiting from address anonymity of the Bitcoin network have attracted many activities such as payments, investments, gambling, and even money laundering in the past decade. 
Unfortunately, some criminal behaviors which took advantage of this platform were not identified. This has discouraged many governments to support cryptocurrency.
Thus, the capability to identify criminal addresses becomes an important issue in the cryptocurrency network. In this paper,  we propose new features in addition to those commonly used in the literature to build a classification model for detecting abnormality of Bitcoin network addresses. 
These features include various high orders of moments of transaction time (represented by block height) which summarizes the transaction history in an efficient way. The extracted features are trained by supervised machine learning methods on a labeling category data set. The experimental evaluation shows that these features have improved the performance of Bitcoin address classification significantly. We evaluate the results under eight classifiers and achieve the highest Micro-F1 / Macro-F1 of 87\% / 86\% with LightGBM.
\end{abstract}


\section{Introduction} \label{sec:introduction}

Since Bitcoin was released in 2008~\cite{nakamoto2008bitcoin}, it has captivated the world with its autonomy and decentralization. Bitcoin is designed as a digital currency system based on peer-to-peer networks instead of a central administration like banks. The proof-of-work protocol allows participants to reach consensus over the distributed network. In addition, all transactions are verified by full nodes and stored in blocks which are chained together by associating previous block header hash. In addition, each block holds the Merkle root of its transactions, which is a kind of fingerprint of transactions, in order to prevent evildoers from tampering data on the blockchain. The properties of immutability, decentralization, data integrity, security of Bitcoin make itself a trustworthy digital currency.

\begin{figure}[t]
\centering
\hbox{
  \hspace{-0.9em}
  \includegraphics[width=1.02\linewidth]{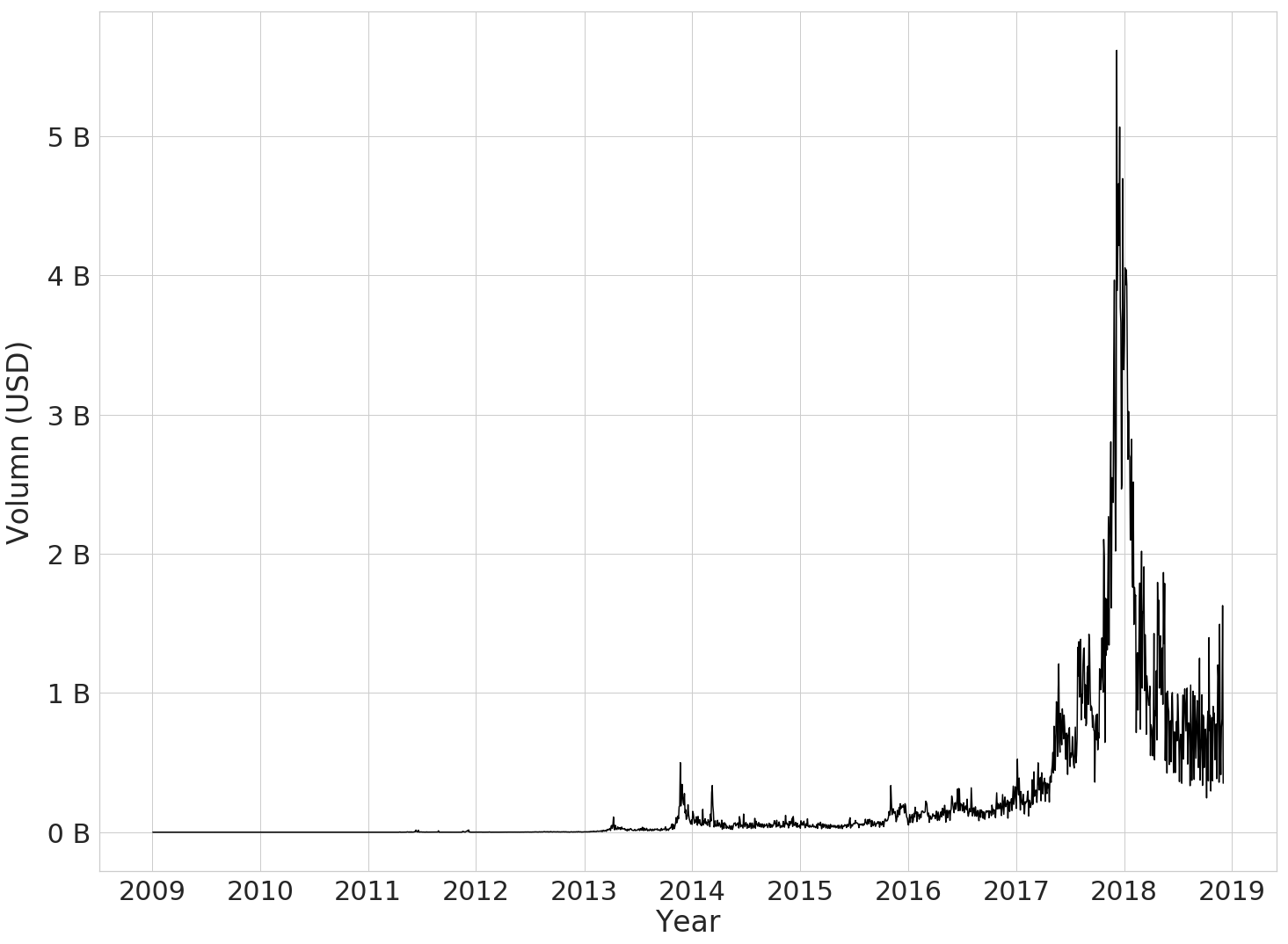}
}
\caption{Daily Transaction Volume of Bitcoin in USD. The estimated value of transactions in USD are retrieved from \textit{Blockchain.com}\setcounter{footnote}{0}\protect\footnotemark . Note that the change of each transaction is excluded.}
\label{fig:bitcoin_volume}
\end{figure}

As the pioneer of thousands of cryptocurrencies, Bitcoin is the most valuable one in terms of market capitalization (market cap). \cite{coinmarketcap} reports that Bitcoin has a market cap of around 59 billion USD, dominating over half of the total market cap of all cryptocurrencies. Moreover, the transaction volume per day on the Bitcoin network in Figure~\ref{fig:bitcoin_volume} raised to billions of US dollars since the mid of 2017 and even once bumped up to 5 billion USD per day. The profitable potential of Bitcoin has attracted people to engage in various activities on Bitcoin, such as payment, investment, gambling, and even laundering.

Some criminal behaviors, such as laundering and frauds, are encouraged by Bitcoin's anonymity. Although Bitcoin is usually described as an anonymous currency, it is actually pseudo-anonymous~\cite{conti2018survey}, i.e, it is hard to link a user to an address by exploring transactions on the blockchain. However, some signs implying the link between addresses and users or between addresses and their usages can be observed. Therefore, there exist messages to possibly identify the criminal behaviors from benign ones. 

\setcounter{footnote}{1}\footnotetext{\url{https://www.blockchain.com/charts/estimated-transaction-volume-usd?timespan=all}}

In this paper, we propose using the \textit{transaction moments} and a series of \textit{extra statistics} as strong features of \textit{transaction history summary}  to identify abnormal addresses. We evaluate the proposed features with eight classifiers using 10-fold cross-validation. We only use the relevant transactions, which is defined as transaction history, of an address or an entity. It is more efficient because the size of data to traverse for a graph pattern might exponentially grow while that of transaction history is only proportional to the relevant transactions of an address or an entity. We finally achieve a Micro-F1 score of 87\% and a Macro-F1 of 86\% with LightGBM~\cite{ke2017lightgbm} in the address-based scheme, which is comparable to the results of prior works~\cite{toyoda2018multi, jourdan2018characterizing}.


This paper is organized as follows: in Section~\ref{sec:bitcoin_network}, we explain Bitcoin network. Next, we investigate several related research on identification, or de-anonymization, of Bitcoin addresses in Section~\ref{sec:related_work}. Section~\ref{sec:proposed_method} illustrates our proposed method: transaction history summary. Data collection, feature extraction, classification, and other training details are elaborated in Section~\ref{sec:experiments}. We then evaluate the classification result in Section~\ref{sec:evaluation_and_discussion} and conclude the paper with Section~\ref{sec:conclusion}.

\section{Bitcoin Network} \label{sec:bitcoin_network}

In the Bitcoin network, transactions specify the number of bitcoins, which are the currency in Bitcoin, to be taken from one address and the number to be transferred to another address. Each transaction can hold multiple inputs and multiple outputs as long as the total amount of inputs is greater than or equal to that of outputs. When making a payment, a user signs the transaction with his private key so as to prove his ownership of the bitcoins to be spent. 

The common unit of Bitcoin is bitcoin (BTC) while each bitcoin is divisible to the eighth decimal place. A BTC can be split into 100,000,000 units, called satoshis, which are the smallest unit of Bitcoin. Most transactions contain transaction fees, which will be transferred to the miner's address as rewards for their proof-of-works. Although transaction fee is not obliged, transactions without any transaction fee or with a lower fee than usual are less likely to be packed to a block by miners. The identities in Bitcoin are private keys. Each private key generates a public key and an address used to public identification. Anyone with a private key is able to spend all bitcoins corresponding to its address. On the other hand, a user can hold an arbitrary number of private keys so it is hard to link an address to a person.

\begin{figure}[t]
\centering
\includegraphics[width=0.45\textwidth]{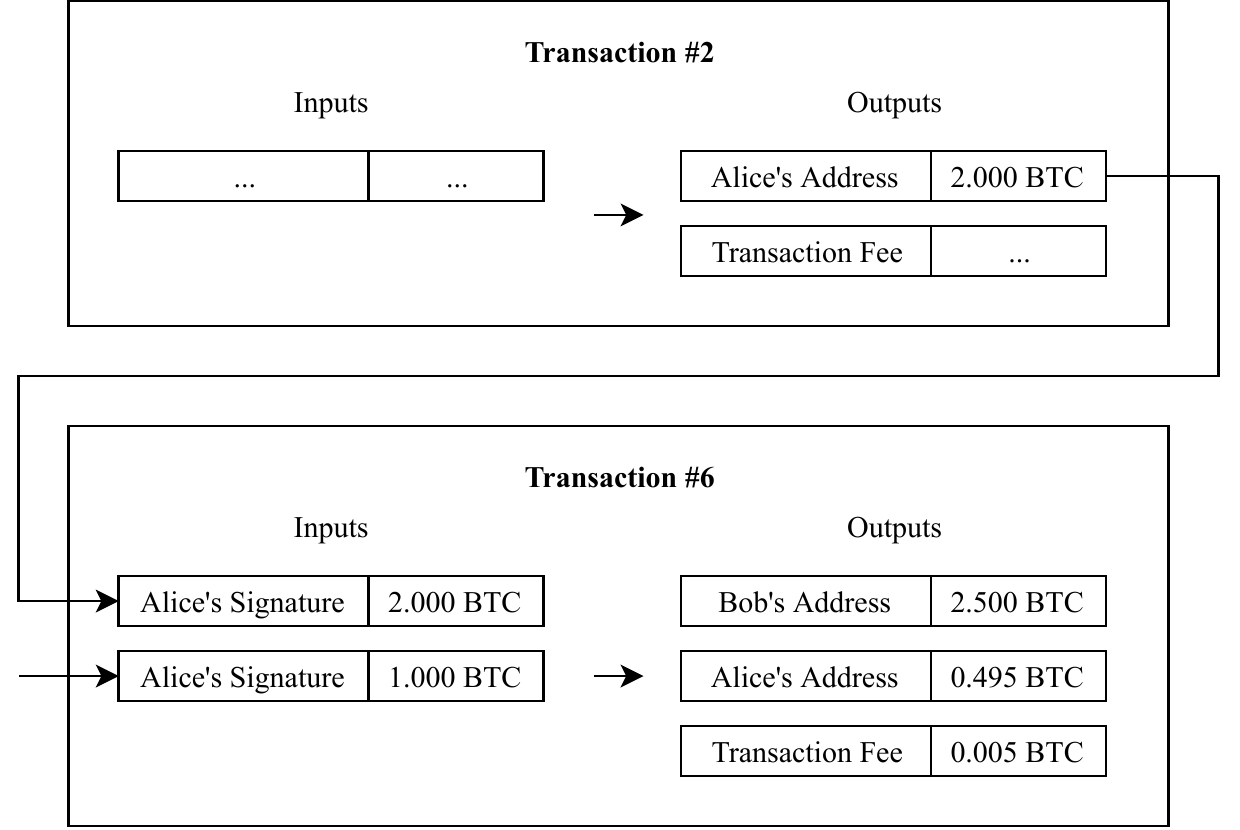}
\caption{An example of Bitcoin transactions.}
\label{fig:bitcoin_transaction}
\end{figure}

Figure~\ref{fig:bitcoin_transaction} shows an example of a transaction that specifies a payment with payback from Alice to Bob. Alice wants to send 2.5 BTC to Bob. She first gathers two of her UTXOs (unspent transaction outputs) holding 2 BTC and 1 BTC, which are somehow received from other addresses. Then Alice signs the two inputs and specifies the outputs as 2.5 BTC to Bob,  0.005 BTC as the transaction fee, and 0.495 BTC back to herself. After the transaction is confirmed on the blockchain, Bob is able to spend the 2.5 BTC.

The Bitcoin network can be viewed as a large composition of transactions. Each transaction is composed of one or multiple inputs and one or multiple outputs. It also records other information such as generation time of blocks. Therefore, we can analyze the whole Bitcoin networks by traversing all the blocks and extract useful information from them.





\section{Related Work} \label{sec:related_work}

There are numerous publications that aimed at Bitcoin network analysis, entity identification, and address de-anonymization. Unlike the descendants emerging in recent years, e.g. Ethereum~\cite{wood2014ethereum}, EOS~\cite{grigg2017eos}, etc., which support smart contract programming, Bitcoin works as a pure transaction ledger. The simplicity makes Bitcoin easy to analyze since the network only contains a bunch of transactions. While data on Bitcoin is open to everyone, some experiments additionally leverage off-chain information such as address tags, indicating potential owners or possible usages of them. These supplementary data put analyzing the Bitcoin network with supervised learning methods into practice.

An intuitive investigation on Bitcoin transactions is to study its transaction flow.  \cite{ron2013quantitative} studied the characteristics of transaction graphs and clustered addresses which might belong to the same entity. Then it is possible to compact a transaction graph into an entity graph. On the other hand, the corresponding input and output addresses confounded by mixing services can be partially understood by transaction graph analysis \cite{moser2013anonymity}.

Several address clustering heuristics~\cite{androulaki2013evaluating, meiklejohn2013fistful} are proposed to link addresses to entities, which represent groups of addresses owned by the same people or the same organizations. The partial linkability between addresses and entities is revealed by several characteristics on the Bitcoin network. Although there is currently no way to link an arbitrary address to its user in the real world, the associated entities are able to be evaluated with off-chain information such as tags (mining pool, exchange wallet, etc.).

A number of studies focused on clustering addresses by unsupervised learning methods. \cite{zambre2013analysis, monamo2016unsupervised, patil2018bitcoin} extract features and cluster address based on statistics of address and patterns of transaction flow in order to detect fraudulent activity in Bitcoin transactions. In these works, k-means~\cite{macqueen1967some} and its variants are adopted to classify features extracted from Bitcoin addresses.

Other studies solve the entity identification problem by supervised learning methods. \cite{yin2017first} classifies cybercriminal entities by supervised learning methods on collected labeled Bitcoin addresses. \cite{bartoletti2018data} train classifiers to detect Ponzi schemes in Bitcoin. To deal with imbalanced data, sampling-based approach and cost-sensitive approach are considered simultaneously in \cite{bartoletti2018data}. To reduce anonymity of Bitcoin by predicting yet-unidentified addresses, \cite{harlev2018breaking} trained classifiers with synthetic minority over-sampling technique~\cite{chawla2002smote} on imbalanced data. 

\cite{ranshous2017exchange} introduces the idea of \textit{motifs} in directed hypergraphs, defining exchange patterns of addresses. In~\cite{jourdan2018characterizing}, the graph-based features \textit{motifs} are then combined with address features, entity features, temporal features, and centrality features to identify Bitcoin entity categories.

Besides graph patterns, transaction history can also provide information for address identification. In \cite{toyoda2017identification}, a set of features are proposed to summarize the transaction history and to identify addresses associated with HYIP based on supervised learning algorithms. These features are extended to identify seven types of Bitcoin-enabled services~\cite{toyoda2018multi}.
 

\section{Proposed Method} \label{sec:proposed_method}

Our work stands on~\cite{toyoda2018multi}, which aimed to identify Bitcoin-enabled service categories based on transaction history summary. We notice the effectiveness of transaction history summary and proposed a series of features to elevate it in various aspects.

A \textit{transaction history summary} is derived from the transaction history, considering merely the direct relevant transactions of a given address or entity. Since the related transactions to an address or an entity contain many redundant fields which cause overhead on the classification model, we extract features from the transactions as transaction history summary. We wish to demonstrate the effectiveness of different types of features. Accordingly, we propose the following three feature types:
\begin{itemize}
  \item Basic Statistics referred from \cite{toyoda2018multi} as the baseline features for address classification, 
  \item Extra Statistics replenishing the address statistics, 
  \item Moments corresponding to transaction distributions.
\end{itemize}


The following subsections provide more details on each type of features.

\subsection{Basic Statistics}

The originally proposed features~\cite{toyoda2018multi} are composed of eight statistical characteristics. Basic statistics include the number of transaction per day, the ratio of received, coinbase and payback transactions to all transactions, the frequencies of different orders of magnitude of transferred bitcoins in spent transactions and in received transactions, and the average numbers of inputs and outputs in the spent transactions. These features are counted numbers divided by duration or a total number. As a result, they characterize transaction history in the aspect of frequency. In \cite{toyoda2018multi}, accuracies of 70\% and 72\% are achieved by these basic statistical features using a random forest classifier in the address-based scheme and the entity-based scheme.

\subsection{Extra Statistics}

Yet there are some characteristics not captured by the basic statistics. We complement the features with extra statistics. The active duration of a series of transactions is defined as $\mathit{lifetime}$, which is the difference between the date of the earliest transaction and that of the latest transaction in terms of the number of days. Mixers tend to have short $\mathit{lifetimes}$ because they are usually disposed of after use. The total received bitcoins and spent bitcoins are taken into consideration, denoted as $\mathit{BTC}_\mathrm{received}$ and $\mathit{BTC}_\mathrm{spent}$. Likewise, the total received and spent money in US dollars are involved as $\mathit{USD}_\mathrm{received}$ and $\mathit{USD}_\mathrm{spent}$. The original values and equivalent values in the real world are both considered in this way. Note that here we count US dollars by converting the amount in each transaction according to its rate at that time. 

In addition to the active duration and total money statistics, the numbers of all types of transactions are included as $n_\mathrm{TX}$, $n_\mathrm{spent}$, $n_\mathrm{received}$, $n_\mathrm{coinbase}$, and $n_\mathrm{payback}$. 
Furthermore, the balances after each transaction are also helpful. We calculate the mean and standard deviation of the balances in BTC and USD to be the last four features, which are $\mu_\mathrm{balance\_btc}$, $\sigma_\mathrm{balance\_btc}$, $\mu_\mathrm{balance\_usd}$, and $\sigma_\mathrm{balance\_usd}$. The reason is that, for example, a mixer might send a large number of bitcoins after it receives them, so its balances would have a large standard deviation than addresses of other categories have.

\subsection{Transaction Moments}

Addresses of different categories have different time distributions of transactions. However, there are no temporal features in the aforementioned statistical features. In order to capture the temporal information, we propose using \textit{Transaction Moments} to encode temporal information as features.

A moment is a quantitative measure of a distribution function. Table~\ref{table:moments} illustrates the distribution behaviors measured by moments of different orders. Generally, four orders are commonly used to describe the shape of a distribution. Therefore, we define first moment (mean), second central moment (variance), third standardized moment (skewness), and fourth standardized moment (kurtosis) of transaction distributions as \textit{Transaction Moments} ($m_n$).

\begin{table}[t]
\captionsetup{justification=centering, labelsep=newline}
\caption{Moments.}
\centering
\begin{tabular}{cc}
    \toprule
    Name & Meaning \\
    \midrule
    $1^{st}\;moment$ & measure of location \\
    $2^{nd}\;moment$ & measure of spread \\
    $3^{rd}\;moment$ & measure of symmetry \\
    $4^{th}\;moment$ & measure of peakedness \\
    \bottomrule
\end{tabular}
\label{table:moments}
\vspace{-1em}
\end{table}

\begin{defn}\label{def:1}
Moment. The n-th moment $\mu_n$ is defined on a real-valued continuous function $f$ as 
\begin{equation}\nonumber
\mu_n = \int_{-\infty}^{\infty} (x-c)^n f(x) dx
\end{equation}
, where $c$ is a central constant. $c$ is usually zero while central moment uses $c$ as the mean of $x$.
\end{defn}

\begin{defn}\label{def:2}
Moment of a continuous random variable. If $f$ is a probability density function and $F$ is its cumulative probability distribution function, the n-th moment is
\begin{equation}\nonumber
\mu_n = \mathrm{E}[{X^n}] = \int_{-\infty}^{\infty} x^n dF(x)
\end{equation}
, where $X$ is a continuous random variable with probability density function $f$.
\end{defn}

\begin{defn}\label{def:3}
Expected value of a discrete random variable.  Let $X$ be a discrete random variable with support $R_X$ and probability mass function $p_X$. The expected value of $X$ is 
\begin{equation}\nonumber
\mathrm{E}[X] = \sum_{x \in R_X} x p_X(x)
\end{equation}
\end{defn}

\begin{defn}\label{def:4}
Moment of a discrete random variable. If $p_X$ is a probability mass function of a discrete random variable $X$, the n-th moment is derived from definition~\ref{def:2} and~\ref{def:3} as 
\begin{equation}\nonumber
\mu_n = \mathrm{E}[{X^n}] = \sum_{x \in R_X} x^n p_X(x)
\end{equation}
\end{defn}

Given the above definitions, we consequently characterize the occurrence time of transactions of an address or an entity as a discrete random variable, which has a probability mass function. Note that the occurrence time here is actually the height of block the transaction subsumed. The mathematical forms of the Transaction Moments we adopt are defined as follows.

\begin{figure*}[!t]
\centering
\includegraphics[width=1.0\textwidth]{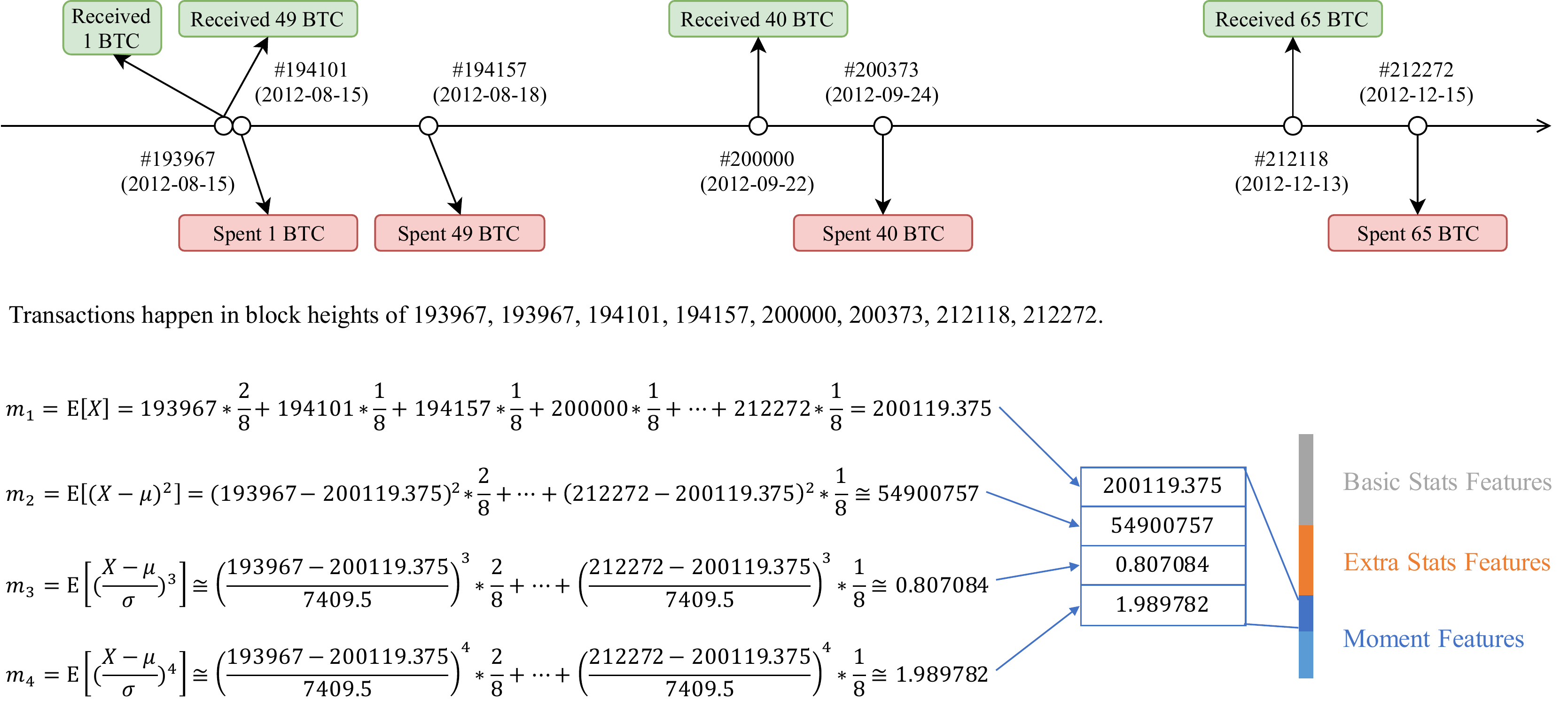}
\caption{An example of transaction history. The upper timeline demonstrates the transaction history of address \textit{15L23mj1TnFa9trXdpQ83iXrGzVd1byKUG}. It is available on public blockchain explorer such as \textit{Blockchain.com}\setcounter{footnote}{1}\protect\footnotemark and \textit{Blockcypher.com}\setcounter{footnote}{2}\protect\footnotemark. The history contains transactions recorded in the \nth{193967}, \nth{194101}, \nth{194157}, \nth{200000}, \nth{200373}, \nth{212118}, \nth{212272} blocks. The overall transaction moments are calculated based on the listed transaction history.}
\label{fig:moments_of_transaction_history}
\end{figure*}
\setcounter{footnote}{2}\footnotetext{\url{https://www.blockchain.com/explorer}}
\setcounter{footnote}{3}\footnotetext{\url{https://live.blockcypher.com}}

\subsubsection{First Moment}

The first moment is identical with mean.

\begin{equation}\label{eq:1}
m_1 = \mathrm{E}[X]
\end{equation}

\subsubsection{Second Central Moment}

We take the second "central" moment as our second moment feature, which is also known as variance. 

\begin{equation}\label{eq:2}
m_2 = \mathrm{E}[(X-\mu)^2],
\end{equation}
where $\mu$ is the expected value of $X$, i.e., $\mu=m_1$.
\vspace{0.5em}

\subsubsection{Third Standardized Moment}

The third moment is centralized and standardized as the term says. 

\begin{equation}\label{eq:3}
m_3 = \mathrm{E}[(\dfrac{X-\mu}{\sigma})^3],
\end{equation}
where $\mu$ is the expected value of $X$ and $\sigma$ is the standard deviation of $X$, i.e., $\mu=m_1$ and $\sigma=\sqrt{m_2}$.
\vspace{0.5em}

\subsubsection{Fourth Standardized Moment}

The fourth moment is also centralized and standardized. 

\begin{equation}\label{eq:4}
m_4 = \mathrm{E}[(\dfrac{X-\mu}{\sigma})^4],
\end{equation}
where $\mu$ is the expected value of $X$ and $\sigma$ is the standard deviation of $X$, i.e., $\mu=m_1$ and $\sigma=\sqrt{m_2}$.
\vspace{0.5em}


Figure~\ref{fig:moments_of_transaction_history} depicts an example of transaction history and demonstrates how to calculate the overall transaction moments of a given address. In this paper, we measure six transaction moments in total. They are moments of \textit{overall transactions}, \textit{coinbase transactions}, \textit{spent transactions}, \textit{reveived transactions}, \textit{payback transactions} as well as \textit{transaction intervals} which are the intervals between transactions in chronological order in terms of block heights. The moments are then concatenated with basic statistics and extra statistics to serve as the transaction history summary.

\section{Experiments} \label{sec:experiments}

To evaluate how effective the proposed features are, we design an experiment of Bitcoin category classification based on addresses and entities. Firstly, we collected labeled data of address-label pairs and fetched all transactions associated with the addresses. The addresses and entities are then be summarized into features with the use of these data. We trained eight supervised classifiers on the extracted features and evaluate the results by average Micro-F1 scores and average Macro-F1 scores of 10-fold cross-validation.

\subsection{Collect Data}

\begin{table}[t]
\captionsetup{justification=centering, labelsep=newline}
\caption{Dataset Details.}
\centering
\begin{tabular}{lccc}
    \toprule
    Category & \# of Entities & \# of Addresses & \# of TXs \\
    \midrule
    Exchange & 158 & 10,466 & 5,701,261 \\
    Faucet & 61 & 340 & 181,602 \\
    Gambling & 90 & 6,733 & 6,536,088 \\
    HYIP & 956 & 2,026 & 377,084 \\
    Market & 18 & 1,900 & 93,930 \\
    Mixer & 32 & 3,199 & 49,064 \\
    Pool & 38 & 1,644 & 274,168 \\
    \midrule
    Total & 1,353 & 26,308 & 13,084,546 \\
    \bottomrule
\end{tabular}
\label{table:dataset_details}
\vspace{-1em}
\end{table}

We leverage the dataset collected by \cite{toyoda2018multi} in order to facilitate the comparison between our method and the previous work. As described in Table~\ref{table:dataset_details}, the dataset contains totally 26,313 addresses with labels and owners, which are derived from a simple heuristic, naming \textit{multi-input transactions}~\cite{androulaki2013evaluating}, \textit{shared-send clustering}~\cite{meiklejohn2013fistful}, \textit{co-spend clustering}~\cite{harlev2018breaking}, or \textit{common spending}~\cite{jourdan2018characterizing}. The idea is that the addresses of inputs in a transaction belong to the same entity because spending bitcoins needs the signature of the owner's private key.


There are 7 categories in total while the data are imbalanced in both address-based scheme and entity-based scheme. We collected the relevant transactions from 2009-01-03 to 2018-06-30 of the addresses, which are over 13 million transactions in total. Some invalid or undecodable transactions are filtered out and addresses/entities containing zero valid transactions are also removed in advance.

\subsection{Summarize Transaction Histories}

We summarize all transaction histories by address and by entity respectively. The selected features are listed in Table~\ref{table:features}, which is divided into three parts: basic statistical features, extra statistical features, and moment features. Their dimensions are 26, 14, and 24 respectively while summed up to be 64 dimensions.

\begin{table}[t]
\captionsetup{justification=centering, labelsep=newline}
\caption{The List of Summarized Features from Transaction History.}
\centering
\settowidth\tymin{$f_\mathrm{received}(10^i)$}
\begin{tabulary}{\linewidth}{CL}
    \toprule
    Feature & Description \\
    \midrule
    $f_\mathrm{TX}$ & The frequency of transactions, defined as number of all transactions per day in the address/entity's lifetime. \\
    $r_\mathrm{received}$ &  The ratio of received transactions to all transactions. \\
    $r_\mathrm{coinbase}$ & The ratio of coinbase transactions to all transactions. \\
    $f_\mathrm{spent}(10^i)$ & The frequency of digit $i$ in USD appeared in spent transactions, where $i\in(10^{-3}, 10^{-2}, \ldots, 10^6)$. \\
    $f_\mathrm{received}(10^i)$ & The frequency of digit $i$ in USD appeared in received transactions, where $i\in(10^{-3}, 10^{-2}, \ldots, 10^6)$. \\
    $r_\mathrm{payback}$ & Payback ratio defined as the ratio of Bitcoin addresses that appear in both inputs and outputs. \\
    $\bar{N}_\mathrm{inputs}$ & The mean value of the number of inputs in the spent transactions. \\
    $\bar{N}_\mathrm{outputs}$ & The mean value of the number of outputs in the spent transactions. \\
    \multicolumn{2}{c}{\rule{120pt}{0.5pt}} \\
    \multicolumn{2}{c}{Basic Statistics} \\
    \midrule
    {$\mathit{lifetime}$} & The duration between the first transaction and the last transaction in terms of days. \\
    {${\mathit{BTC}_\mathrm{spent}}$} & Total spent BTC. \\
    {${\mathit{BTC}_\mathrm{received}}$} & Total received BTC. \\
    {${\mathit{USD}_\mathrm{spent}}$} & Total spent USD, which are converted based on daily BTC/USD rates from \textit{Coinmarketcap.com}~\cite{coinmarketcap}. \\
    {${\mathit{USD}_\mathrm{received}}$} & Total received USD, which are converted based on daily BTC/USD rates from \textit{Coinmarketcap.com}~\cite{coinmarketcap}. \\
    $n_\mathrm{TX}$ &  The number of transactions. \\
    $n_\mathrm{spent}$ & The number of spent transactions. \\
    $n_\mathrm{received}$ &  The number of received transactions. \\
    $n_\mathrm{coinbase}$ & The number of coinbase transactions. \\
    $n_\mathrm{payback}$ &  The number of payback transactions. \\
    {$\mu_\mathrm{balance\_btc}$} & The mean value of balance in BTC after each transaction. \\
    {$\sigma_\mathrm{balance\_btc}$} & The standard deviation of balance in BTC after each transaction. \\
    {$\mu_\mathrm{balance\_usd}$} & The mean value of balance in USD after each transaction. \\
    {$\sigma_\mathrm{balance\_usd}$} & The standard deviation of balance in USD after each transaction. \\
    \multicolumn{2}{c}{\rule{120pt}{0.5pt}} \\
    \multicolumn{2}{c}{Extra Statistics} \\
    \midrule
    {$\mathit{m}_\mathrm{n,overall}$} & The moments of overall transaction distribution. \\
    {$\mathit{m}_\mathrm{n,spent}$} & The moments of spent transaction distribution. \\
    {$\mathit{m}_\mathrm{n,received}$} & The moments of received transaction distribution. \\
    {$\mathit{m}_\mathrm{n,coinbase}$} & The moments of coinbase transaction distribution. \\
    {$\mathit{m}_\mathrm{n,payback}$} & The moments of payback transaction distribution. \\
    {$\mathit{m}_\mathrm{n,interval}$} & The moments of transaction interval distribution. \\
    \multicolumn{2}{c}{\rule{120pt}{0.5pt}} \\
    \multicolumn{2}{c}{Moments} \\
    \bottomrule
\end{tabulary}
\label{table:features}
\vspace{-1em}
\end{table}

The three types of coinbase, spent, received are mutually exclusive. Specifically, a transaction is assigned as a coinbase transaction if it contains a coinbase input. If a transaction has no coinbase input and the address appears in some of its inputs, the transaction is identified as a spent transaction. Otherwise, if the address appears only in the outputs, it is assigned as a received transaction.

With regard to the moment features we propose, we measure the moments of overall transaction distribution, spent transaction distribution, received transaction distribution, coinbase transaction distribution, payback transaction distribution, and transaction interval distribution. We deem that how often an address or an entity has transactions is important in order to reveal what category it is. In addition, not only the frequency, moments of transaction distributions and transaction interval distributions can characterize behaviors such as an address active in the beginning but listless in recent days, and even periodic behaviors. Whether a transaction relates to spending Bitcoins, receiving Bitcoins, or even spending Bitcoins back to the spender matters. They are also indications of the usages of their addresses.

In~\cite{toyoda2018multi}, only up to 1000 successive transactions of an entity are summarized due to the huge data size. We instead collected all related transactions of the addresses in the dataset so as to extract features from all transactions of an address or an entity. The whole history is characterized into the extracted transaction history summary.

\subsection{Train Classifiers}


The extracted features, or transaction history summaries of an address, are then classified by machine-learning-based algorithms. We would like to compare our features to recent studies in Bitcoin address classification. Most of them classified the extracted features with several machine learning methods. Therefore, in this work, we evaluate them with eight classifiers: Logistic Regression, Perceptron~\cite{rosenblatt1958perceptron}, Support Vector Machine (SVM)~\cite{hearst1998support}, Adaptive Boosting with Decision Tree (AdaBoost)~\cite{freund1997decision, schapire1999brief, hastie2009multi}, Random Forest (RF)~\cite{breiman2001random}, Extreme Gradient Boosting (XGBoost)~\cite{chen2016xgboost}, Light Gradient Boosting Machine (LightGBM)~\cite{ke2017lightgbm}, and Neural Network~\cite{haykin2004comprehensive}.

\subsection{Implementation Details}

For the first seven classifiers, we exploit the Python machine learning library - Scikit-learn~\cite{pedregosa2011scikit}. Each classifier is tuned by grid search with 10-fold cross-validation in order to find a good set of parameters respectively. The decision tree-based methods are not affected by data normalization. We simply normalize the features with division by the maximum absolute value of each dimension for the classifiers that are not based on decision trees (Logistic Regression, Perceptron, and SVM).

The neural networks are implemented with Keras~\cite{chollet2015keras}. The architecture is composed of four fully-connected layers of hidden size 512 with batch normalization and dropout regularization. The model is followed by an output fully-connected layer of size 7 at the end. 
Similarly, we normalize the data in advance since neural networks are sensitive to input data. 


As can be seen in Table~\ref{table:dataset_details}, 
the category imbalance exists in both schemes. 
To deal with the imbalance, we adopt the stratified random sampling, split training set and validating set, and take up the cost-sensitive learning to train classifiers. The weight of each sample is calculated by

\begin{equation}\label{eq:6}
    w_i = k / p_i
\end{equation}
, where $k$ is a constant usually defined as $\frac{1}{\mathit{\#\,of\,categories}}$ and $p_i$ is the probability of the category of the sample.
\vspace{0.5em}

\begin{figure}[t]
  \centering
  \includegraphics[width=1.0\linewidth]{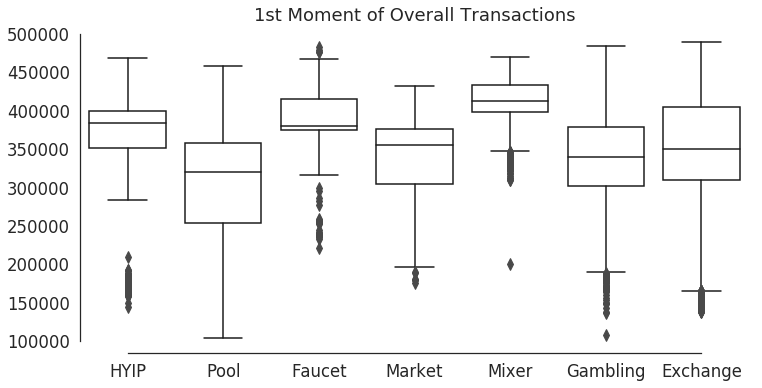}
  \caption{The first moment of overall transactions of addresses from seven categories. The upper, middle, lower lines on the boxes represent the quartiles and the highest datum and lowest datum are 1.5 IQR from the Q1 and from the Q3. The data points out of whiskers are plotted as dots.}
  \label{fig:1st_moment}
  \vspace{-.5em}
\end{figure}

Another pitfall is the bias in transaction temporal distribution. We observe a bias among different categories caused by data collection. In Figure~\ref{fig:1st_moment}, it is obvious that some categories like Gambling, Faucet, and Pool are distinguishable merely by their first moments of the distributions. Therefore, we normalize the distributions of overall transactions, coinbase transactions, spent transactions, received transactions, and payback transactions, while excluding transaction intervals. In practice, we subtract  random variables by their minimum value as follows.

\begin{equation}\label{eq:7}
X' = X - min(X),
\end{equation}
where $min(X)$ is the minimum value in $X$.
\vspace{0.5em}


The first moment of Equation~\ref{eq:1} is replaced by

\begin{equation}\label{eq:8}
m_1' = \mathrm{E}[X'] = \mathrm{E}[X] - min(X)
\end{equation}
, which we call the first min-shifted moment of $X$.
\vspace{0.5em}

Other moments are \textbf{central} moments so they are not affected by the min-shift. For the example in Figure~\ref{fig:moments_of_transaction_history}, the first min-shifted moment is computed as $m_1 = 200119.375 - 193967 = 6152.375$.

\section{Evaluation and Discussion} \label{sec:evaluation_and_discussion}

To show the effectiveness of the proposed methods, we evaluate on (i) Micro-F1 scores and Macro-F1 scores, (ii) selected features, (iii) the confusion matrix, and (iv) important features. For (i), we compare the results among eight supervised machine learning classification algorithms on the transaction history summaries of the labeled addresses and entities. Then the same metrics (accuracy, precision, and F1-score) are used to evaluate the features (ii) that we select by an ablation study on different feature combinations. 

\begin{table}[t]
\captionsetup{justification=centering, labelsep=newline}
\caption{Results of Supervised Classifiers with Full Features.}
\centering
\begin{tabular}{l|cccccc}
    \toprule
    Method & \multicolumn{2}{c}{Entity-based Scheme} & \multicolumn{2}{c}{Address-based Scheme} \\
     & Micro-F1 & Macro-F1 & Micro-F1 & Macro-F1 \\
    \midrule
    Logistic Regression & 0.73 & 0.60 & 0.48 & 0.45 \\
    Perceptron & 0.69 & 0.53 & 0.36 & 0.35 \\
    SVM & 0.56 & 0.46 & 0.47 & 0.46 \\
    AdaBoost & 0.30 & 0.30 & 0.36 & 0.36 \\
    Random Forest & 0.90 & 0.73 & 0.83 & 0.81 \\
    XGBoost & 0.90 & 0.77 & 0.83 & 0.82 \\
    LightGBM & 0.90 & 0.75 & \textbf{0.87} & \textbf{0.86} \\
    Neural Network & \textbf{0.91} & \textbf{0.78} & 0.83 & 0.81 \\
    \bottomrule
\end{tabular}
\label{table:result_different_model}
\vspace{-1em}
\end{table}

The best Micro-F1 scores are 91\% for the entity-based scheme and 87\% for the address-based scheme. However, the Macro-F1 scores are 78\% and 86\% instead.
Table~\ref{table:result_different_model} shows the detailed results of two schemes with Micro-F1 scores and Macro-F1 scores. 
A Micro-F1 computes the overall average F1 score over the testing set, whereas a Macro-F1 computes the F1 score independently for each category and take the average, treating all categories equally. We suspect that the large difference between entity-based F1 scores suffers from the data imbalance and data scarcity. For the entity-based scheme, Market has only 1 or 2 samples in the testing when evaluating by 10-fold cross-validation. Consequently, we focus on the address-based scheme in further experimental evaluations.

The Random Forest, XGBoost, LightGBM, and Neural Network are the best four machine learning methods among the eight classifiers in our experiment. Comparing to the prior work~\cite{toyoda2018multi}, that has achieved an accuracy of 72\% in the entity-based scheme and 70\% in the address-based scheme, we achieve a better result by Random Forest with the proposed features. In general, we showed that the categories are identified more accurately by classifiers working with our proposed features.

Although the best result in entity-based is achieved by neural networks, LightGBM performs  most stably in both schemes. As a result, we use the best model of LightGBM to illustrate (iii) confusion matrix and (iv) important features in the following part. On the other hand, with moments and extra statistics, as seen in the table, the decision tree-based classifier and perceptron work better on address-scheme while neural network, SVM and logistic regression have better results on the entity-based scheme.

\begin{table}[b]
\captionsetup{justification=centering, labelsep=newline}
\caption{Ablation Study of Basic Statistics, Extra Statistics and Moments in Address-based Scheme.}
\centering
\begin{tabular}{cccccc}
    \toprule
    \multicolumn{3}{c}{Method} & \multicolumn{2}{c}{Results} \\
    B & E & M & Micro-F1 & Macro-F1 \\
    \midrule
    $\surd$ &  &  & 0.79 & 0.78 \\
     & $\surd$ &  & 0.77 & 0.74 \\
     &  & $\surd$ & 0.66 & 0.57 \\
    $\surd$ & $\surd$ &  & 0.86 & 0.85 \\
    $\surd$ &  & $\surd$ & 0.84 & 0.82 \\
    $\surd$ & $\surd$ & $\surd$ & \textbf{0.87} & \textbf{0.86} \\
    \bottomrule
\end{tabular}
\label{table:result_ablation}
\vspace{-1em}
\end{table}

 The performance of LightGBM with different combinations of features is presented in Table~\ref{table:result_ablation}. The features are divided into three types: Basic Statistics, Extra Statistics and Moments. 
 Our result shows the effectiveness of feature combinations. Evaluated with any combinations of features, the result is better than evaluated with any single feature.
It also implies that the basic statistics and extra statistics capture two main behaviors, whereas moments remedy the transaction history summary. Although moments alone do not work as well as statistical features, moments boost up the F1 scores when combined with other features. 
We achieve the best result with basic statistics, extra statistics, and moments mixed together.

\begin{figure}[t]
    \vspace{-1em}
    \centering
    \includegraphics[width=1.0\linewidth,trim=0 0 10 10,clip]{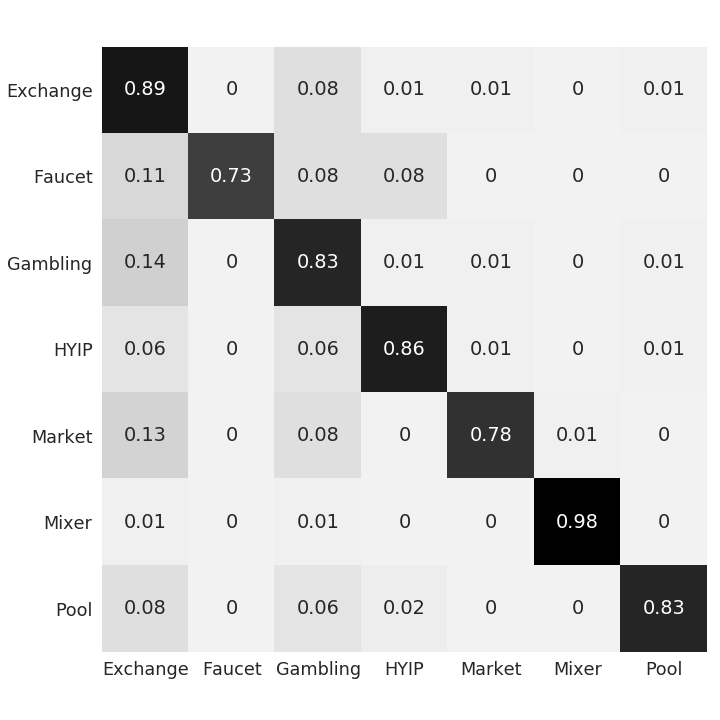}
    \vspace{-2em}
    \caption{The confusion matrix of LightGBM trained with all features in the address-based scheme. The values in grids are categorical accuracy, where each row is supposed to be summed up as 1. Besides, the darkness of each grid is proportional to its value.}
    \label{fig:confustion_matrix}
    \vspace{-1em}
\end{figure}

The confusion matrix of LightGBM with all features in the address-based scheme is depicted in Figure~\ref{fig:confustion_matrix}. Most categories are classified well, indicating that the classifier works well on these categories. The classification accuracy of Mixer is even 96\% and the second high one is 84\% of Exchange. Other categories achieve at least 70\% in terms of accurate classification.

\begin{table}[b!]
\captionsetup{justification=centering, labelsep=newline}
\caption{Top 20 Important Features According to the Model.}
\centering
\subfloat[%
  The top 10 features.
]{
\begin{minipage}[t]{0.45\linewidth}
\centering
\begin{tabular}{lr}
    \toprule
    Feature Name & Feature Type \\
    \midrule
    $f_{TX}$ & Basic Stats \\
    $\bar{N}_\mathrm{outputs}$ & Basic Stats \\
    $\bar{N}_\mathrm{inputs}$ & Basic Stats \\
    $n_\mathrm{received}$ & Extra Stats \\
    $m_\mathrm{1,interval}$ & Moments \\
    $\sigma_\mathrm{balance\_btc}$ & Extra Stats \\
    $\mathit{lifetime}$ & Extra Stats \\
    $r_\mathrm{payback}$ & Basic Stats \\
    $m_\mathrm{1,received}$ & Moments \\
    $\mu_\mathrm{balance\_btc}$ & Extra Stats \\
    \bottomrule
\end{tabular}
\end{minipage}
}
\hfill
\subfloat[%
  The \nth{11} to \nth{20} features.
]{
\begin{minipage}[t]{0.45\linewidth}
\centering
\begin{tabular}{lr}
    \toprule
    Feature Name & Feature Type \\
    \midrule
    $f_{received}$ & Basic Stats \\
    $n_\mathrm{spent}$ & Extra Stats \\
    $n_\mathrm{TX}$ & Extra Stats \\
    $m_\mathrm{2,received}$ & Moments \\
    ${\mathit{BTC}_\mathrm{spent}}$ & Extra Stats \\
    $\mu_\mathrm{balance\_usd}$ & Extra Stats \\
    $m_\mathrm{1,total}$ & Moments \\
    $m_\mathrm{2,total}$ & Moments \\
    ${\mathit{BTC}_\mathrm{received}}$ & Extra Stats \\
    $f_\mathrm{received}(10^2)$ & Basic Stats \\
    \bottomrule
\end{tabular}
\end{minipage}
}
\label{table:top_20_features}
\vspace{-2.2em}
\end{table}

Table~\ref{table:top_20_features} illustrates the 20 most important features in LightGBM classifier, which achieves best result in the address-based scheme. The features are sorted by the information gain importance~\cite{louppe2013understanding} from largest to smallest. As can be seen from the table, six out of the ten features are proposed by us. Although the moment features are less important than the other two kinds, they take one-fourth of the top twenty features. The extra statistics, on the other hand, appear almost the same times as the basic statistics. For more details about important features, please refer to Appendix~\ref{sec:feature_importance}.


\section{Conclusion} \label{sec:conclusion}

In this work, we introduce new features as transaction history summary for Bitcoin address and entity classification. 
The transaction history summary is composed of basic statistics, extra statistics, and transaction moments. The basic statistics are based on the previous work~\cite{toyoda2018multi} and capture the features in the aspect of frequency. The extra statistics additionally contain total amounts and statistical measures of transactions. The transaction moments characterize the temporal distribution of transactions as well as transaction intervals.

Our experiment showcases the performance benefits from using our proposed features for Bitcoin address/entity classification. The combinations of features make huge progress in terms of classification accuracy. Moreover, our proposed features dominate the ten most important features according to a well-trained LightGBM classifier. As the best result we achieve, the Micro / Macro F1 scores are 87\% / 86\% in the address-based scheme. The high accuracy in each category is indicated from measuring the similarity between Micro-F1 and Macro-F1. Also, the confusion matrix of our best result further proves it. The entity-based classification, however, suffers from data imbalance and data scarcity. Therefore, we plan to do the experiment on a larger dataset \cite{jourdan2018characterizing} in the future work so as to evaluate the entity-based scheme.


\newpage

\appendices

\section{Feature Importance}
\label{sec:feature_importance}

\begin{figure}[h]
\hspace{-14pt}
\scalebox{1.0}[0.81]{
\includegraphics[width=0.5\textwidth,trim=144 12 144 7,clip]{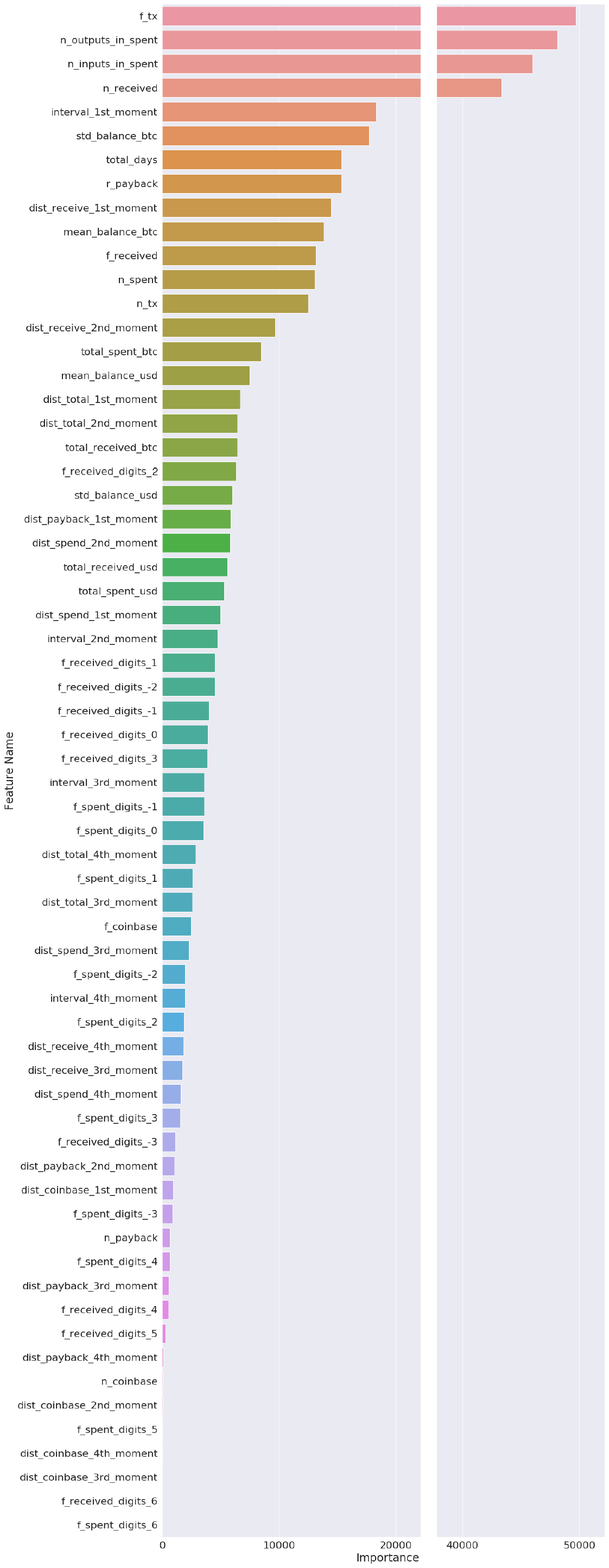}
}
\caption{Feature importance scores are reported as the total information gains of splits for each feature in LightGBM.}
\label{fig:feature_importance}
\end{figure}

\newpage

\end{document}